\documentclass[aps,prl,preprint,superscriptaddress]{revtex4-2}

\usepackage{graphicx}

\begin{document}


\title{Study of hydrogen absorption in a novel three-dimensional graphene structure: Towards hydrogen storage applications}



\author{Aureliano Macili}
\affiliation{NEST, Istituto Nanoscienze-CNR and Scuola Normale Superiore, Piazza S. Silvestro 12, 56127 Pisa, Italy}
\author{Ylea Vlamidis}
\affiliation{NEST, Istituto Nanoscienze-CNR and Scuola Normale Superiore, Piazza S. Silvestro 12, 56127 Pisa, Italy}
\affiliation{Department of Physical Science, Earth, and Environment, University of Siena, Via Roma 56, 53100, Siena, Italy}
\author{Georg Pfusterschmied}
\author{Markus Leitgeb}
\author{Ulrich Schmid}
\affiliation{Institute of Sensor and Actuator Systems, TU Wien, 1040, Vienna, Austria}
\author{Stefan Heun}
\affiliation{NEST, Istituto Nanoscienze-CNR and Scuola Normale Superiore, Piazza S. Silvestro 12, 56127 Pisa, Italy}
\author{Stefano Veronesi}
\email[]{stefano.veronesi@nano.cnr.it}
\affiliation{NEST, Istituto Nanoscienze-CNR and Scuola Normale Superiore, Piazza S. Silvestro 12, 56127 Pisa, Italy}



\date{\today}

\begin{abstract}
The use of a novel three-dimensional graphene structure allows circumventing the limitations of the two-dimensional nature of graphene and its application in hydrogen absorption. Here we investigate hydrogen-bonding on monolayer graphene conformally grown via the epitaxial growth method on the ($0001$) face of a porousified $4$H-SiC wafer. Hydrogen absorption is studied via Thermal Desorption Spectroscopy (TDS), exposing the samples to either atomic (D) or molecular (D$_2$) deuterium. The graphene growth temperature, hydrogen exposure temperature, and the morphology of the structure are investigated and related to their effect on hydrogen absorption.
The three-dimensional graphene structures chemically bind atomic deuterium when exposed to D$_2$. This is the first report of such an event in unfunctionalized graphene-based materials and implies the presence of a catalytic splitting mechanism. It is further shown that the three-dimensional dendritic structure of the porous material temporarily retains the desorbed molecules and causes delayed emission. The capability of chemisorbing atoms after a catalytic splitting of hydrogen, coupled to its large surface-to-volume ratio, make these structures a promising substrate for hydrogen storage devices.
\end{abstract}


\maketitle

\section{Introduction}
In view of reducing fossil fuel consumption, global interests are shifting towards the use of renewable energy sources. However, renewables present a series of issues \cite{Heal_2009} which limit their widespread applicability.

The intermittency of renewable sources could be managed by employing a form of chemical storage of the renewable energy \cite{Ager_2018}. When adopting this solution, hydrogen is considered one of the most suitable forms of chemical storage, since it can be produced by electrolysis of water, and its oxidation only involves the emission of H$_2$O, thus forming a completely sustainable cycle.

Practical hydrogen-energy applications, however, require a technology allowing the reliable storage of large hydrogen quantities. Compressed gas cylinders and cryogenic liquefied hydrogen are used, but present practical, economical, and safety issues \cite{Eberle2009,Verfondern2008}. The most promising form of hydrogen storage is the use of solid-state solutions, which are envisioned to allow physical or chemical storage in much more favourable pressure and temperature conditions \cite{Rusman_2016}.

The use of graphene in solid state hydrogen storage applications is addressed in recent studies  \cite{Bonfanti_2018,Hornek_r_2006a,Mohan2019} and results particularly promising due to a series of advantageous physico-chemical properties of the material \cite{Tozzini2013} such as the relative ease of transition from $sp^2$ to $sp^3$ hybridization due to curvature \cite{Tozzini_2011}. Moreover, graphene is lightweight, inert, and can be functionalized \cite{Mashoff_2013, Torres2020}, all desirable properties for a storage material. However, due to its two-dimensional nature, hydrogen storage devices based on pristine flat graphene would require prohibitively large surface areas. One solution to this issue is the development of a three-dimensional structure that retains the outstanding properties of graphene, thus allowing the folding of a large graphene surface in a compact structure. We refer to such a system as "$3$D graphene".

The 3D graphene samples used in this work are obtained via epitaxial graphene growth \cite{Yazdi_2016} on the ($0001$) surface of $4$H-SiC (Si-face) wafers that have been previously porousified with photoelectrochemical etching, resulting in an extremely large surface area. Such substrates are known as porous silicon carbide \cite{Leitgeb_2017a,Leitgeb_2016,Leitgeb2017,Leitgeb_2018} and have been already successfully used in  MEMS applications \cite{Leitgeb_2017}. The graphene is obtained using the well-known epitaxial growth method \cite{RiedlPhd}. This method consists in the thermal decomposition of a SiC wafer under ultra-high vacuum conditions. The heating causes the sublimation of Si and Si-compounds, leading to an accumulation of C atoms on the surface. These atoms re-arrange in the hexagonal geometry typical of graphene. The first layer that forms is known as \textit{buffer layer}, which maintains $30 \%$ of its C atoms covalently bound to Si atoms \cite{Lauffer2008}. Further sublimation of Si leads to the growth of a second buffer layer which, as it develops, detaches the former one rendering it a layer of free standing monolayer graphene, bound to the substrate only by van der Waals interactions \cite{Mishra2016}. The graphene layers grown with this method closely retrace the morphology of the $3$D substrate. They coat the pores of the SiC structure and thus produce a meandering, three-dimensional structure. The resulting material presents a surface $\sim 200$ times larger than that of a same-dimensions flat graphene sample, as reported in a recent publication by S. Veronesi et al. \cite{Veronesi_2022}.

Here we study the hydrogen absorption properties of this novel three-di\-men\-sio\-nal graphene substrate. All samples utilized in the present work were obtained following the growth procedure described in \cite{Veronesi_2022} for the Si-face porous 4H-SiC. The samples were characterized by STM and Raman spectroscopy. TDS of the $3$D sample, and its comparison with flat epitaxial graphene (see the SI for details), confirmed the detection of above room-temperature chemisorption signals after exposure of the sample to molecular deuterium. This effect was never before reported in unfunctionalized graphene samples and implies the catalytic splitting of hydrogen on the sample surface. The coupling of hydrogen splitting and hydrogen storage on 3D graphene offers promising prospects for the use of this material in future hydrogen storage devices.

The emission of deuterium molecules physisorbed on this 3D graphene was also investigated and resulted in a "Delayed Emission Model", which describes the release of molecules from the sample. Specific measurement procedures were designed to test this model and allowed the study of the effect of the porous three-dimensional structure on the emission of desorbed molecules.

\section{3D Graphene Characterization}

\begin{figure}
   \includegraphics[width=\linewidth]{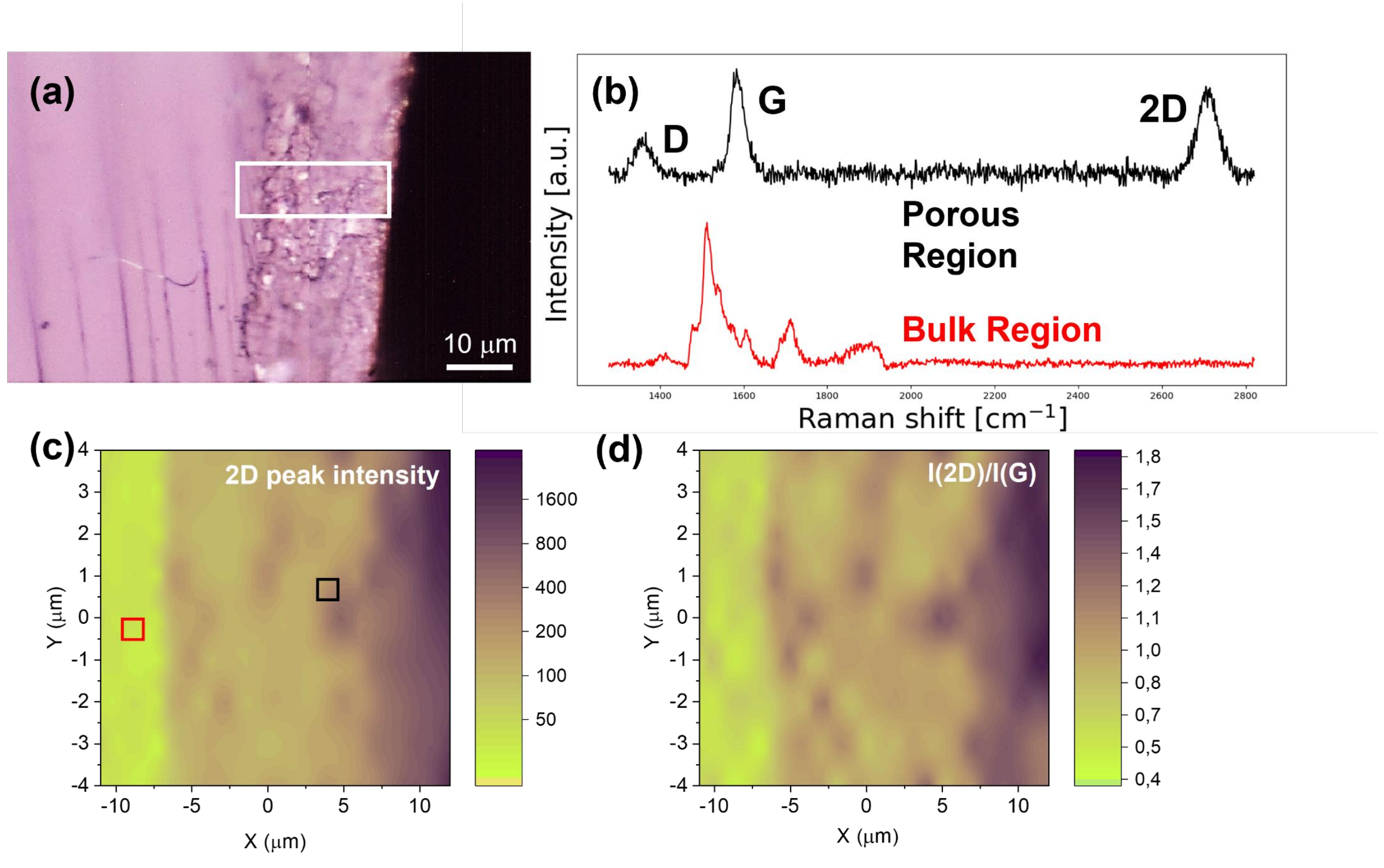}
   \caption{\label{cleaved}Raman analysis of the sample in cross-section after graphene growth. (a) Optical Microscopy image of the cross-section, (b) extracted Raman spectra of marked regions in (c) (red and black box). (c) Raman mapping of 2D band intensity and (d) intensity ratio of 2D/G band.}
\end{figure}

The surfaces of the 3D graphene samples used in this work were characterized by Raman spectroscopy (Figure S$1$ in the SI) which indicates a homogeneous coverage with mainly monolayer graphene. The Raman benchmarks reported in the SI are fully consistent with results obtained in \cite{Veronesi_2022}. In addition, the structures were also investigated via STM measurements (Figure S$2$ in the SI), allowing the visualization of the porous structure, as well as of the graphene monolayer and buffer layer covering the flatter portions. Lastly, one of the samples used in the hydrogenation experiments was cleaved, allowing the measurement of a Raman spectrum in the exposed inner porous structure, reported in Figure \ref{cleaved}, and SEM analysis of the pores. The heat maps of Raman benchmarks for graphene reported in Figure \ref{cleaved} demonstrate the presence of graphene in the inner regions of the porousified sample. This confirms that the porous structure is uniformly covered by few layer graphene which grows conformally to the complex SiC topography.

\section{Results and Discussion}

To clarify the experimental conditions of the Thermal Desorption Spectroscopy (TDS) measurements, four prefix will be utilized in the following. In particular, D (D$_2$) labels a TDS experiment on a sample exposed to atomic deuterium (molecular deuterium). Moreover, RT or LT labels indicate TDS experiments on samples hydrogenated at room temperature, about 30~$^\circ$C, or at low temperature, about $-160$~$^\circ$C. Further details of the hydrogenation procedure are provided in the Methods section.

\subsection{TDS After Room Temperature Hydrogenation (RT TDS)}

\begin{figure}
   \includegraphics[width=\linewidth]{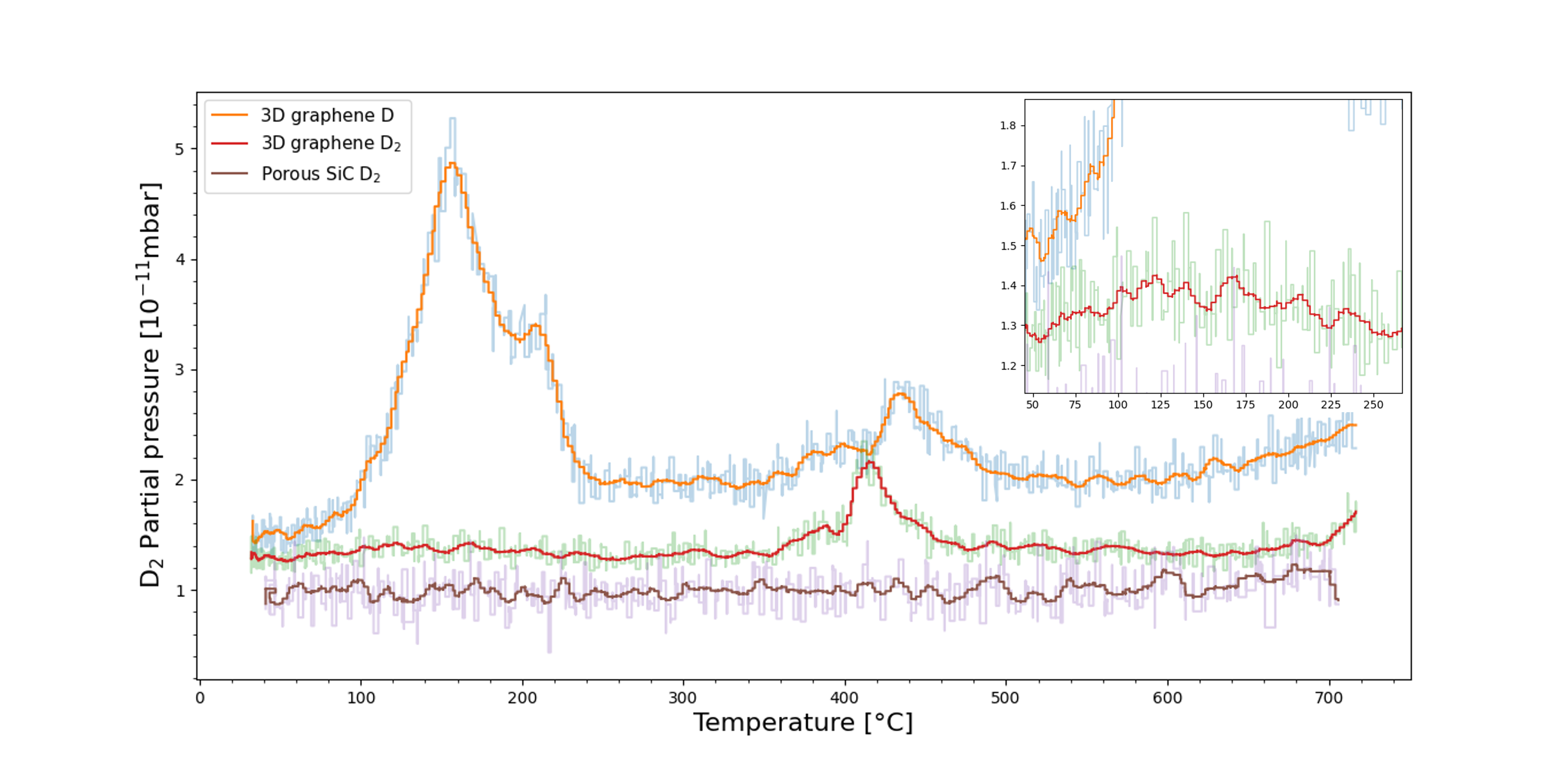}
   \caption{RT TDS spectra of 3D graphene samples after D hydrogenation (orange) or D$_2$ hydrogenation (red). The RT D$_2$ TDS spectrum of a pristine porous SiC sample is also reported (purple). The inset reports the RT D$_2$ TDS spectrum, plotted individually, to better appreciate the low intensity peak observed in the lower temperature region. Lines represent the TDS spectra after a smoothing process. The raw TDS data are shown in the background of every plot.}
   \label{RTTDS}
\end{figure}

 Hydrogen storage in this novel 3D graphene material was studied using TDS (see Methods section for details). TDS spectra were acquired for a sample of porous SiC graphenized following the procedure described in \cite{Veronesi_2022} (see the Methods section for details). Such samples are simply referred to as "3D graphene" in the following. Figure \ref{RTTDS} reports the RT TDS spectra of a 3D graphene sample after exposure to D and D$_2$, in addition to the RT TDS spectrum of a pristine porous SiC sample (without graphene) after D$_2$ hydrogenation.

The RT D TDS spectrum of 3D graphene samples features two distinct signals. The first signal is a prominent and broad peak, centered at around 155~$^\circ$C, corresponding to a 1.09 eV binding energy (see the Methods section for details on the binding energy calculation). It presents a high temperature shoulder consistently observed at 210~$^\circ$C. The second signal is weaker and was consistently observed at 430~$^\circ$C (1.86 eV binding energy). Additionally, in the RT D TDS spectra, we observe a linear background signal which increases with temperature, while the other spectra present a flat background.

The RT D$_2$ TDS spectra of the 3D graphene sample also feature two distinct emissions. The first signal consists of a broad, weak peak, centered at around 130~$^\circ$C (1.01 eV, shown in the inset to Figure \ref{RTTDS}). The second peak is sharp, more intense, and centered at around 410~$^\circ$C (1.81 eV).

No desorption signal is detected in the RT TDS measurement of pristine porous SiC, regardless of the hydrogenation procedure to which it was subjected (D or D$_2$). 

The presence of desorption signals observed in RT TDS spectra of the 3D graphene samples after exposition to atomic or molecular hydrogen, together with the absence of similar signals in the TDS spectrum of the untreated porous SiC sample, demonstrates that the presence of graphene is a mandatory requirement for the hydrogen adsorption process to take place.

The absence of peaks in the TDS spectra of pristine porous SiC allows to exclude that the peaks observed in the other two spectra of Figure \ref{RTTDS} are caused by hydrogen interaction with some contaminant, possibly left by the etching process, or by hydrogen adsorption on the sample holder or manipulator.

The temperatures at which the signals appear in the RT TDS spectra of $3$D graphene are imputable to chemisorption, since it is the only form of stable interaction with the sample at temperatures exceeding 30~$^\circ$C \cite{_ljivan_anin_2009,Miura_2003,Zecho_2002}. This implies that all signals observed in the RT TDS spectra are the product of desorption of deuterium atoms chemically bonded to the graphene. However, deuterium molecules (D$_2$) are electronically stable and unable to form covalent bonds without previously breaking the D-D bond. Therefore, the observation of a chemisorption peak implies the formation of a covalent bond between a single deuterium atom and graphene. This observation suggests the existence of a catalytic hydrogen-splitting mechanism that
allows molecular deuterium to divide into its individual atoms which are then able to interact
with the surface by chemisorption. The existence of such a catalytic mechanism is also supported
by the similarity of the signal positions in the RT D TDS and RT D$_2$ TDS spectra. Indeed, the desorption events are registered at similar temperatures regardless
of the hydrogenation procedure, a strong indication that they refer to the
same physical phenomenon, i.e. the desorption of deuterium atoms chemically bound to the sample.

The signal intensity (in particular for the 150~$^\circ$C signal) is very different when comparing the spectra of the RT TDS in the two hydrogenation scenarios. This discrepancy is consistent with the hypothesis of a D$_2$ splitting mechanism. Indeed, in the case of RT D TDS the atoms are directly fed to the sample through the hydrogen cracker. The energy required to break the D$_2$ bond is already provided by the instrument. On the other hand, in the case of RT D$_2$ TDS, the D atom production takes place on the sample and is determined by the splitting reaction. As a consequence, the production rate is limited both kinetically (the reaction must overcome an activation barrier) and numerically (the amount of active sites is limited and sites can saturate).

The different signal intensities between the RT TDS spectra in the two hydrogenation scenarios can, therefore, be explained by the greater amount of D atoms that are able to form a bond in the D hydrogenation case with respect to the D$_2$ hydrogenation case. The similar intensity of the high temperature peaks in the two hydrogenation scenarios can be explained assuming that, in both cases, the chemisorption sites of that interaction get saturated by deuterium atoms in both hydrogenation scenarios, regardless of the difference in number of atoms reaching the sample.

\subsection{TDS of Samples Subjected to Multiple Annealing Steps}

\begin{figure}
   \includegraphics[width=\linewidth]{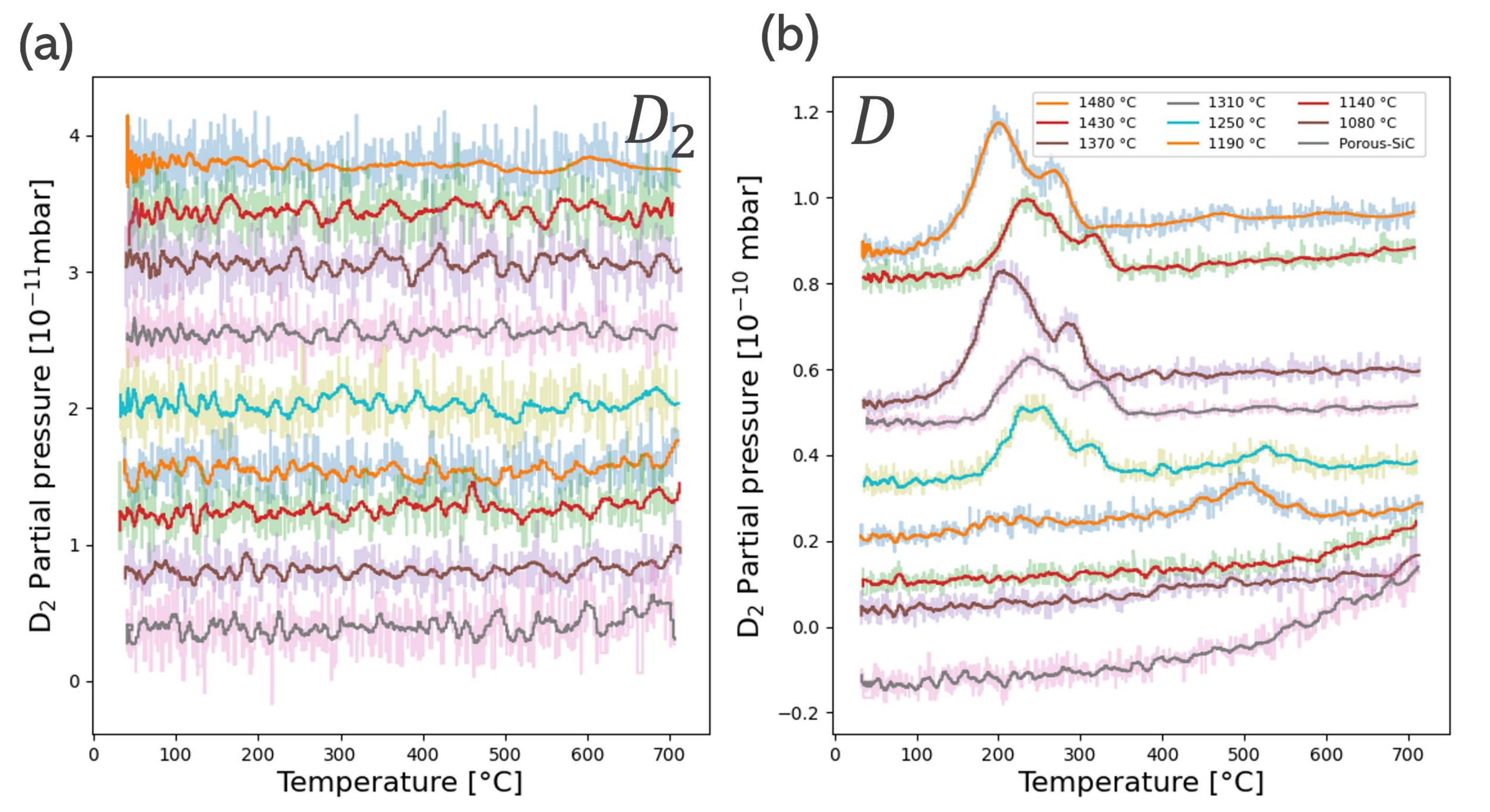}
   \caption{ Two sets of RT TDS spectra of a porous SiC sample subjected to a series of increasing-temperature annealing processes. (a) RT D$_2$ TDS spectra, (b) RT D TDS spectra. In each set, the lowest plot refers to untreated porous SiC, while the successive curves are RT TDS spectra taken after successive annealing steps. The annealing temperatures are indicated in the label. In each set, successive plots are shifted by an increasing offset to avoid overlaps and to facilitate reading.}
  \label{history}
\end{figure}

To better understand the dependence of the RT TDS spectra of the porous SiC material on surface carbonization states ”in between” untreated porous SiC and optimally grown 3D graphene, a porous SiC sample was subjected to a series of subsequent increasing-temperature annealing steps.
Specifically, the sample was subjected to eight consecutive $2$-minutes annealing steps at temperatures ranging
from 1080~$^\circ$C to 1480~$^\circ$C, each time increasing the annealing temperature by 60~$^\circ$C.
After each annealing step, the sample was exposed to molecular or atomic deuterium at room temperature, and
the corresponding RT TDS spectra were measured. Figure \ref{history} reports the series of $8$ RT TDS spectra after the different temperature annealing steps, as well as a first spectrum measured on pristine porous SiC. Figure \ref{history}(a) refers to D$_2$ hydrogenation, while \ref{history}(b) to its atomic counterpart.

Figure \ref{history}(a) clearly shows that the exposure of the sample to molecular deuterium never results in any
form of hydrogen storage, regardless of the annealing temperature (and, therefore, of the surface carbonization state). No signal emerging from the background level is detected in any of the reported
RT D$_2$ TDS spectra.
On the other hand, when considering atomic hydrogenation, the RT D TDS spectra are
largely different. The presence and intensity of the signals depends on the annealing temperature of the sample. This further confirms that the sample holder or possible sample contaminations have no effect on the observed TDS desorption peaks.

For untreated porous SiC, as well as after the first two annealing steps, the RT D TDS spectra do not feature any signals. Once the sample undergoes the third (1190~$^\circ$C) annealing cycle,
the RT D TDS spectrum starts showing a high-temperature peak centred at 480~$^\circ$C. A sample subjected to a 1190~$^\circ$C annealing is expected to present graphene on its surface, as well as a large amount of buffer layer,
formed as a precursor to actual monolayer graphene. When the
sample is subjected to a fourth (1250~$^\circ$C) annealing, its RT D TDS spectrum appears similar to the one reported in Figure \ref{RTTDS} for the RT D TDS spectrum of 3D graphene, in that it shows the presence of two distinct peaks. By further annealing the sample for $2$ minutes at 1310~$^\circ$C, the resulting RT D TDS spectrum only shows a signal at 200~$^\circ$C, while the high temperature
peak previously observed results undetectable. Every successive RT D TDS spectrum, taken after higher temperature annealing steps, results in the presence of only the 200~$^\circ$C peak. In particular, it is relevant to notice the complete absence of a 500~$^\circ$C signal in the RT D TDS spectrum of the multiply-annealed sample after the 1370~$^\circ$C annealing step, differently from what has been observed in the 3D graphene which was grown at this temperature (cf.~Figure \ref{RTTDS}).

The main facts resulting from the TDS measurements on a sample subjected to multiple annealing
processes are the complete absence of a signal in all RT D$_2$ TDS spectra and the variability
of the RT D TDS spectra measured after each annealing step.

The similarity of the RT D TDS spectra of the sample after its fourth annealing (1250~$^\circ$C) and of the RT D TDS of the optimally grown sample (1370~$^\circ$C) suggests that treating a porous SiC sample with multiple
annealing facilitates graphene growth even at temperatures lower than those required with a single-annealing step. This effect could
be the result of a longer (cumulative) heating time ($8$ minutes in total after the fourth annealing compared to a total of $5$ minutes for the optimal growth). It could be caused
by the presence of graphene islands, formed during previous annealing processes, which
act as crystallisation sites, facilitating further graphene development in successive annealing
steps.

The appearance of the 500~$^\circ$C peak in the RT D TDS spectra measured after the third annealing can be explained by attributing that specific signal to the chemisorption of deuterium atoms on the buffer layer \cite{Lin2014}. Indeed, when graphene growth is starting, the amount of buffer layer (which acts as a precursor for monolayer formation) is expected to be maximal. This hypothesis is also consistent with the disappearance of the high temperature signal after the fifth annealing. Indeed, cyclic heating and increasing temperature annealing are expected to facilitate graphene development. This results in the complete covering of all initially exposed buffer layer areas and the disappearance of its relative chemisorption signal.  

The complete absence of signals in any of the RT D$_2$ TDS spectra of the multiple-annealed sample, regardless
of the annealing temperature, can be interpreted as the effect of an hindrance of the
previously discussed catalytic splitting mechanism. Whatever causes the deuterium molecules to catalytically divide into atoms during a D$_2$ hydrogenation is removed by subjecting the
sample to repeated annealing cycles. Crystalline defects are extremely reactive sites that have
been reported to act as centres for hydrogen splitting \cite{Vitiello_2000}. Moreover, defects are expected
to be reduced in number after an annealing procedure. The absence of signals in the RT D$_2$ TDS spectra can thus be explained by considering the catalytic site in our sample to be some form of crystalline defect, either in the graphene or in the porous SiC
substrate. Due to their three-dimensional nature, more crystallographic defects are expected in 3D graphene samples in comparison to graphene grown on a flat SiC surface.

Therefore, a porous SiC sample will initially present a large number of defects and no graphene on its surface. After it undergoes multiple annealing cycles, the number of defects will decrease as the graphene develops. Both graphene and catalytic spitting are required to observe signals in a D$_2$ TDS but, when performing multiple annealing, the two features are in competition and do not coexist. For this reason, no desorption signal can be observed in the spectra.

In conclusion, the results of the RT D$_2$ TDS measurements, performed on samples subjected to multiple annealing steps, indicate that chemisorption can take place after D$_2$ exposure only if the sample presents both graphene \textit{and} surface defects. This contemporary presence cannot be obtained in samples subjected to multiple annealing steps, while it is achieved in optimally grown 3D graphene.
The TDS study of samples subject to multiple annealing steps also serves as a confirmation of the fact that amorphous carbon does not contribute to hydrogen absorption in these samples. Indeed the amount of non-crystalline carbon regions on the sample is expected to increase with successive annealing cycles, but the intensity of the desorption signals does not follow a similar trend.

\subsection{TDS After Low Temperature Hydrogenation (LT TDS)}

In addition to the RT TDS measurements, samples subjected to the optimal growth procedure were also studied in the case in which the hydrogenation was conducted with samples cooled down to $-160$~$^\circ$C. The aim of these LT TDS measurements is to detect the low-energy physisorption interactions \cite{Tozzini_2011,Patchkovskii_2005} between deuterium and 3D graphene.

\begin{figure}
   \includegraphics[width=\linewidth]{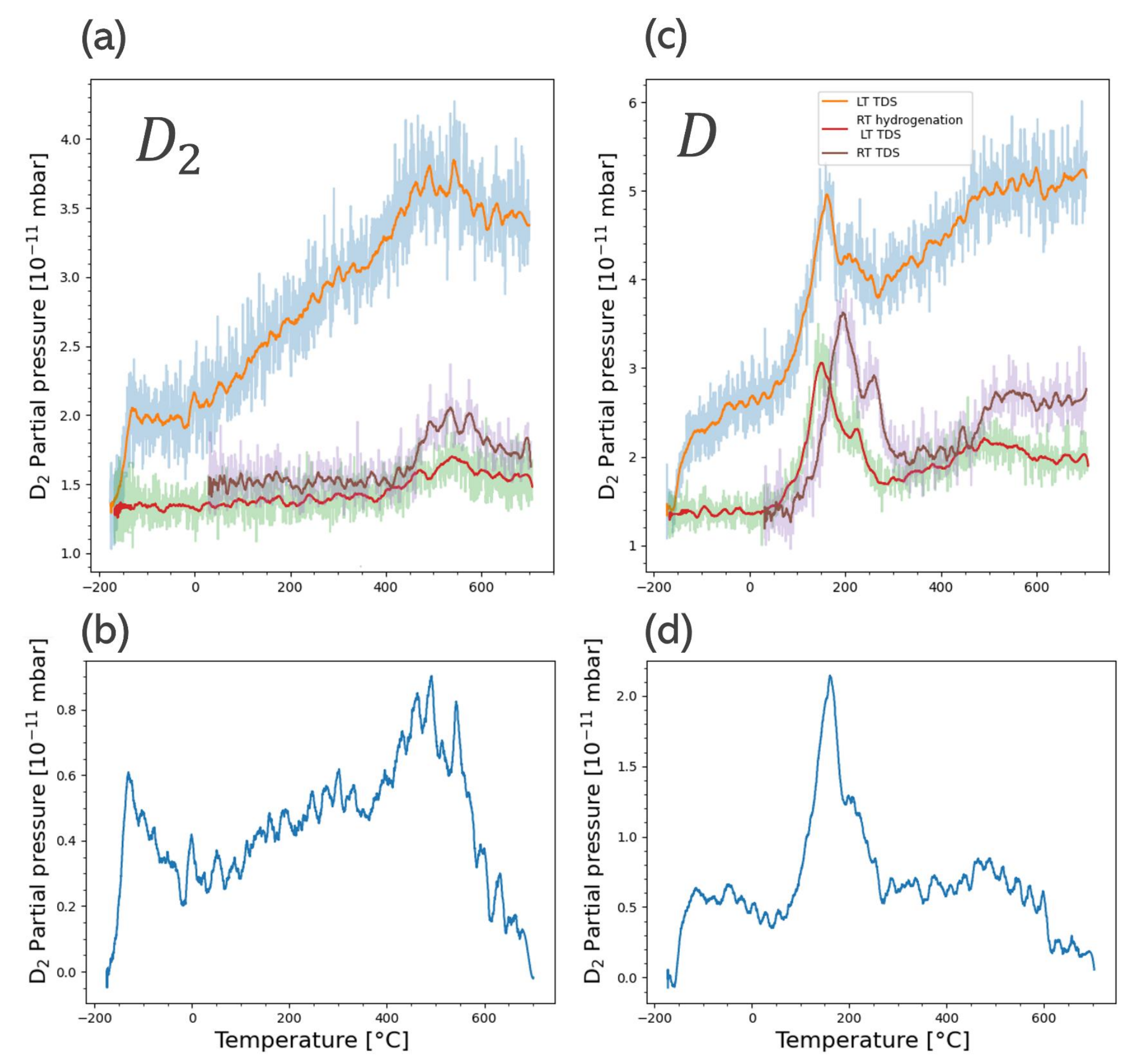}
   \caption{\label{LTTDS}Two sets of TDS measurements performed on 3D graphene. (a) D$_2$ hydrogenation,  (c)  D hydrogenation. In each set, the orange curve refers to a standard LT TDS and the red curve to a RT hydrogenation LT TDS. For comparison, in each set, the RT TDS plot is also reported in purple.  (b) LT D$_2$ TDS spectrum (orange plot in (a)) after subtraction of a linear background, (d) LT D TDS spectrum after background subtraction.}
\end{figure}

Figure \ref{LTTDS}(a) and \ref{LTTDS}(c) report a comparison between LT TDS and RT TDS spectra of a 3D graphene sample in both hydrogenation scenarios. The most striking feature of the LT D$_2$ TDS spectrum is the increase of the background TDS signal with increasing temperature. This trend is substantial and results in a mostly linear background, onto which the desorption peaks are superimposed. To gain a better understanding of the desorption signal positions, the baseline of each LT TDS spectrum was calculated and subtracted. In the case of LT D$_2$ TDS, this allows the identification of three distinct signals (Figure \ref{LTTDS}(b)). The first one consists of a sharp peak centered at $-131$ $^\circ$C (0.36 eV). The second signal is hardly distinguishable in the background noise but can be observed at 160 $^\circ$C, while the third peak has a maximum at 490 $^\circ$C. The second and third peaks are similar in shape, position, and intensity to the peaks observed in the RT D$_2$ TDS spectra and can be interpreted as the product of the chemisorption phenomena previously discussed. The peak at $-131$ $^\circ$C is the product of D$_2$ physisorption. A third spectrum (red curves in Figures \ref{LTTDS}(a) and \ref{LTTDS}(c)) is also measured and labeled RT hydrogenation LT TDS. In this measurement, the 3D graphene samples are subjected to hydrogenation procedure at room-temperature and then cooled to $-160 \ ^\circ$C before starting the TDS measurement.

Similarly to what observed in the case of D$_2$, the desorption spectrum of the cold sample after atomic deuterium deposition features a pronounced linear trend, on top of which three additional signals can be distinguished. The subtraction of the baseline allows a more precise characterisation of these peaks (Figure \ref{LTTDS}(d)). The first signal is centered around $-120$~$^\circ$C (0.40 eV), the second consists of a main peak presenting a maximum at 160~$^\circ$C and a higher temperature shoulder, the third signal is less intense and is centred around 480~$^\circ$C. Also in this case, the two higher temperature signal are similar, both in position and shape, to the chemisorption signals observed in the RT D TDS of the same sample.

Physisorption of atomic hydrogen on carbon structures is considered an exotic phenomenon, only possible at very low temperatures \cite{Bonfanti_2018}. The operative condition of the LT TDS are not so extreme,
therefore the detection of atomic deuterium physisorption can be excluded, leaving D$_2$ physisorption as a possible candidate. The first peaks detected in the LT TDS both for D and D$_2$ hydrogenation can thus be attributed to the desorption of physisorbed deuterium molecules.
In the case of LT D TDS, the presence of D$_2$ can be explained as the product of
D atoms recombination. Attribution of the low temperature signals to physisorption is consistent with literature, since peaks at similar temperature were observed in hydrogenated low-temperature carbon structures \cite{Krkljus_2013,Takahashi_2016}.

The presence of chemisorption peaks in the high temperature ($>30$~$^\circ$C) part
of the LT D$_2$ TDS spectrum indicates the fact that the splitting mechanism leading to the
production of atomic species during D$_2$ exposure remains active even when the sample is
cooled to $-160$~$^\circ$C. With a similar deduction, in relation to the presence of chemisorption
peaks in the LT TDS spectra, it is possible to conclude that none of the proposed chemisorption
interactions presents an activation energy large enough to be hindered by the cooling of the
sample.

Physisorption interaction cannot be observed (or indeed take place) when the sample is exposed to hydrogen while at room temperature, since the thermal energy is too large to allow for van der Waals interactions to be stable. In RT hydrogenation LT TDS measurements, hydrogen deposition takes place at room temperature but the spectrum is measured from $-160$~$^\circ$C. These spectra can, therefore, be interpreted as a standard LT TDS in which all spectral features imputable to hydrogen physisorption have been removed from the plot. This technique was designed to investigate the nature of the linear background signal observed in the LT TDS spectra. 

Indeed, for both hydrogenation scenarios, the LT TDS spectra are drastically different between hydrogenation performed at room temperature or on the cold
sample. In particular, in RT hydrogenation LT TDS measurements, the increasing background trend is strongly reduced and the physisorption peak disappears. This implies that the background increase observed in the standard LT TDS spectra
is a consequence of the desorption of deuterium molecules interacting via physisorption with the cold sample. 
 In addition, the similarity between the RT TDS spectra and the corresponding part of the RT hydrogenation LT TDS spectra in Figure \ref{LTTDS} confirms that in both cases the signal is due to chemisorption.

\subsection{The Delayed Emission Model}

The attribution of the background signal in LT TDS to the desorption of physisorbed deuterium molecules is not sufficient to explain the shape of this contribution to the spectrum. In fact, this signal is present in the entire temperature range of the measurement, and its intensity increases as the
sample temperature rises. Moreover, the existence of this signal in the high temperature ($>30$~$^\circ$C) region of the spectrum is in apparent contrast with its identification as a product of physisorption.

The desorption behaviour observed during standard LT TDS can be rationalised
using a model that we named ”Delayed Emission”. This model is
based on the idea that the three-dimensional nature of the graphene structure causes the desorbed molecular deuterium to be temporarily trapped inside the pores before effusing from the sample.

\begin{figure}
   \includegraphics[width=0.9\linewidth]{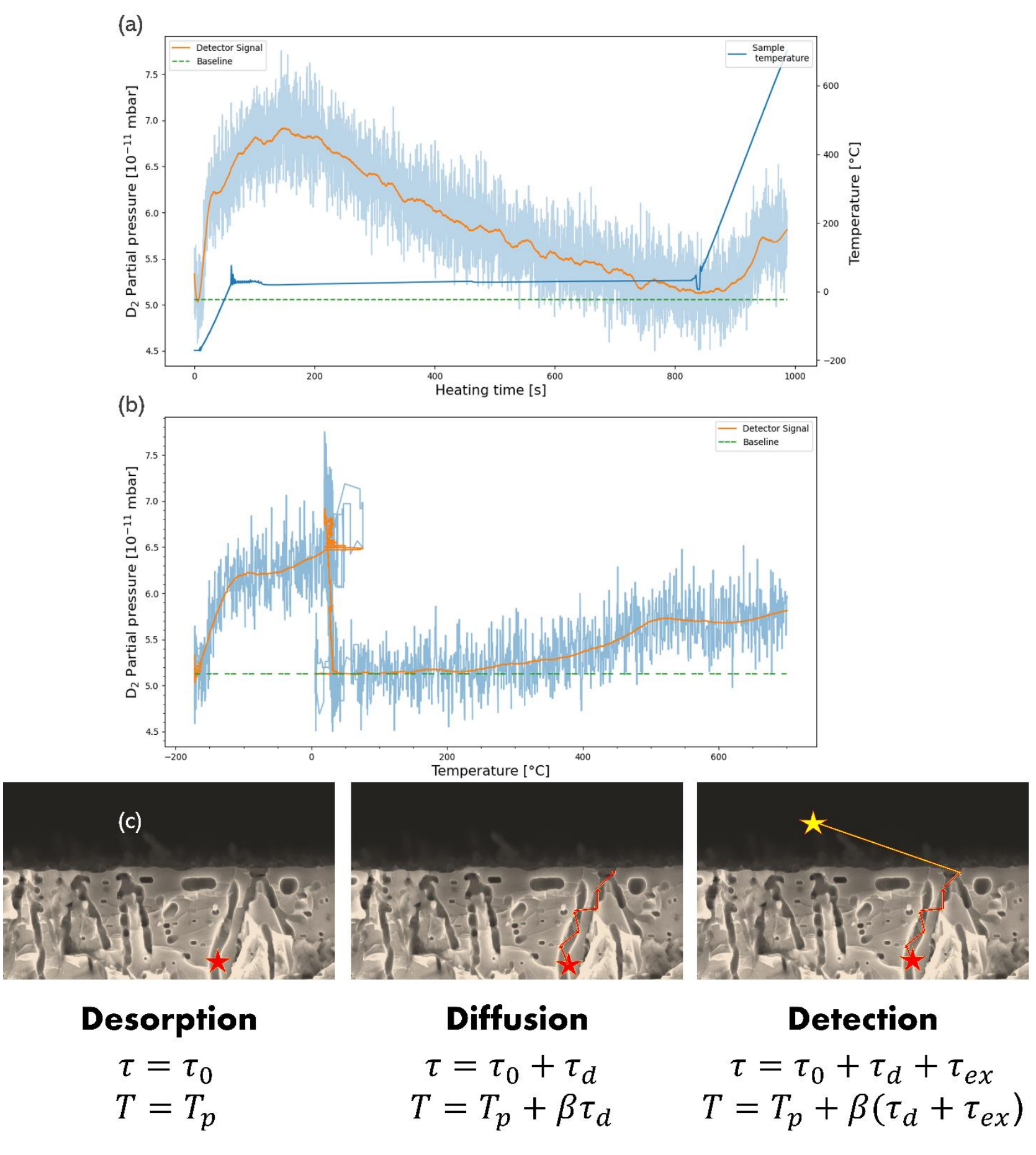}
   \caption{\label{DEM} Results of the modified measurement procedure. (a) the orange graph is the D$_2$ partial pressure as a function of the heating time, $p$D$_2$(t) (signal intensity is reported on the left $y$ axis). The blue line is the sample temperature as a function of the elapsed time (sample temperature is reported on the right $y$ axis). (b) reports the same signal as a function of the sample temperature, $p$D$_2$(T). (c) Schematic representation of the delayed emission model overlayed on a SEM image of the cleaved sample. $T$ refers to temperature, $\tau$ to time. $\beta$=$\frac{\partial T}{\partial \tau}$ is the heating ramp slope, $T_p$ the desorption temperature and $\tau_0$ the desorption time, $\tau_d$ is the time required by a molecule to exit the sample, $\tau_{ex}$ the time to reach the detector. Considering the large mean free path of molecules in UHV conditions, $\tau_{ex}$ is assumed to be negligible with respect to $\tau_{d}$.}
\end{figure}

This results in a time lag between the actual desorption of molecules and their detection by
the mass spectrometer. Thus, the delayed emission of hydrogen from the sample results in a fallacious
attribution of the signal to temperatures higher than the actual desorption temperature.
Figure \ref{DEM}(c) reports a schematic representation of the desorption path leading to delayed emission.
The delayed emission of molecular hydrogen from regions of
the structure located at different depth results in the ”spread” of the relative signal in a larger time window and thus temperature range, producing a background increase instead of an individual peak. In the context of the delayed emission
model, the increasing intensity of the background signal can be rationalized by considering the
fact that the number of ”delayed” deuterium molecules increases as more desorption events
take place. In addition, a higher sample temperature is also likely to result in a shorter ”exit
time” due to an increase in the trapped particle velocity, resulting in an increased number of
molecules being able to exit the sample during the final part of the TDS measurement.

We designed a tailored measurement procedure to test the delayed emission model. It is conceptually similar to a standard LT TDS but uses a different heating program, effectively splitting the standard linear ramp into three sections: (a) Ramp I, (b) Temperature Hold, and (c) Ramp II. During the first phase (Ramp I), the sample is subjected to linear heating ramp, from the starting temperature to 30~$^\circ$C. As the sample reaches room temperature, the heating program is switched to its second phase. Here, the temperature is maintained at 30~$^\circ$C (Temperature Hold). During the temperature hold phase, the sample can be maintained at room temperature for any given time period. In the present investigation, the time interval was of the order of $15$ minutes. The final phase (Ramp II) is identical to the first one and consists in a second linear heating ramp, increasing the sample temperature from room temperature to 700~$^\circ$C.

By allowing the sample to rest for 15 minutes at room temperature during the temperature
hold phase, all molecules temporarily retained within the sample are provided
enough time to find their way out of the structure and reach the detector. If the delayed emission model is accountable,
when the heating is restarted during Ramp II, the background contribution observed
in a standard LT TDS spectrum is expected to be absent or drastically reduced, since (at least
initially) the sample would contain no products of previous desorptions. Figure \ref{DEM}(a) reports the variation in D$_2$ partial pressure as a function of the elapsed time during a modified LT TDS measurement. The modified heating ramp is also reported (sample temperature as a function of elapsed time). Figure \ref{DEM}(b) shows the plot of D$_2$ partial pressure as a function of temperature.

The analysis of the $p$D$_2$(t) plot reported in Figure \ref{DEM}(a) allows to draw of some conclusions
regarding the desorption behaviour in time. In particular, it is shown that, as the sample
reaches room temperature and the heating ramp enters the temperature hold phase, the spectrometer
signal keeps increasing for minutes before reaching a maximum value and starting to decrease.
After peaking, the amount of desorbed D$_2$ decreases steadily, until the partial pressure in the measurement chamber is restored to the initial value (before the start of Ramp
I). As the heating program progresses to Ramp II, the signal intensity maintains fairly steady until a chemisorption peak is detected, resulting in a new signal increase. The re-apperance of a linear background signal after desorption of chemisorbed deuterium atoms is coherent with the delayed emission model, and the same effect can also justify the slight background increase observed in the RT D TDS spectra reported in Figure \ref{RTTDS} and Figure \ref{history}.

The corresponding behaviour in the $p$D$_2$(T) spectrum can be observed in Figure \ref{DEM}(b). During Ramp I, the signal is, as expected, similar to the first part of the LT D$_2$ TDS spectrum
reported in Figure \ref{LTTDS}. During the 30~$^\circ$C temperature hold, the signal intensity rises and falls. The region of the spectrum at temperatures
above 30~$^\circ$C refers to Ramp II. There the signal is distinctively lowered with respect to the value registered at the onset of the temperature
hold and remains close to background level up to 500~$^\circ$C where a chemisorption peak is registered at temperatures compatible with the results reported in Figure \ref{RTTDS}.

The experimental observations reported in Figure \ref{DEM} are in agreement with the proposed delayed emission model. The $p$D$_2$(t) signal increase after the onset
of the temperature hold phase can be interpreted as the delayed detection of the molecules
desorbed during Ramp I. The D$_2$ partial pressure lowering to the initial value during the temperature hold is an indication of the fact that the system eventually stops emitting hydrogen,
i.e. no molecules are retained in the structure. Concerning the $p$D$_2$(T) spectrum, the drastic reduction of the signal after the temperature hold phase is a clear
indication of the fact that the majority of the background signal detected above 30~$^\circ$C, in a standard
LT D$_2$ TDS, is the product of the delayed detection of molecules that detached at lower
temperatures.

D$_2$ TDS measurements performed on flat graphene also confirm the fact that the background signals observed in Figure \ref{LTTDS} are completely imputable to the three-dimensional structure of the 3D graphene samples (See Figures S$4$ and S$5$ in SI).

A comparison between a standard LT D$_2$ TDS and the $p$D$_2$(T) plot obtained with a 30~$^\circ$C temperature hold can be used to obtain a quantitative confirmation of the delayed emission model. In particular, the number of D$_2$ moles retained inside the sample is obviously independent of the heating program followed. Therefore, the only difference between the two measurement procedures is the moment of emission of the trapped molecules. In the measurement featuring a temperature hold phase, the emission takes place during the temperature hold phase, while in a standard LT TDS the delayed emission gets distributed over the entire region of the spectrum above RT.

As a consequence, the number of D$_2$ moles emitted during the temperature hold phase in the modified measurement procedure ($n_{TH}$) should equal the number of moles contributing to the background signal in a standard LT TDS for temperatures $>30$~$^\circ$C, $n_{DE}$. The procedure followed for the calculation of these numbers of molecules is reported in the supporting information (Figures S$6$ and S$7$). The final values for the two quantities are indeed very similar, adding further confirmation to the delayed emission model.

\section{Conclusions}
This work demonstrates hydrogen absorption in a novel three-dimensional graphene structure epitaxially grown on porousified monocrystalline SiC. RT TDS measurements prove the existence of two forms of deuterium chemisorption in 3D graphene, both stable at room temperature.
These peaks were observed exposing the sample to atomic or molecular deuterium. The presence of chemisorption peaks after D$_2$ hydrogenation is particularly interesting since it was never observed before in unfunctionalized graphene and it is an achievement toward the development of practical solid-state hydrogen storage devices. The observation of chemisorption peaks in RT D$_2$ TDS spectra implies the presence of a catalytic hydrogen splitting reaction taking place on the sample surface.

Cyclic heating was found to hinder the chemisorption of hydrogen during D$_2$ hydrogenation, suggesting the catalytic sites might be crystalline defects of the porous SiC substrate or of the graphene structure. Moreover, chemisorption of atomic deuterium was found to depend on the quality of the graphene layer developed on the porous SiC. The higher temperature desorption peak reported in D TDS spectra was attributed to chemisorption of deuterium atoms on the graphene buffer layer. This experiment also leads to the conclusion that cyclic heating promotes graphene growth on SiC.

A low temperature desorption signal was observed in 3D graphene with both hydrogenation procedures. It was attributed to the physisorption of D$_2$ molecules to the graphene surface.
LT TDS spectra also present a substantial linear increase of the background signal during the measurement. This spectral feature was proven to be the product of molecular deuterium physisorption as well. The observation of the increasing background signal was explained by a "Delayed Emission Model" in which the desorbed molecules are temporarily retained in the three-dimensional structure, causing a time lag between their desorption and detection. The delayed emission model was confirmed by introducing a modified measurement procedure which allows the entrapped D$_2$ molecules to exit the structure while the sample temperature is maintained constant. The model was also confirmed quantitatively, by showing that the amount of deuterium molecules exiting the sample during the temperature hold phase equals the amount of deuterium molecules contributing to the background increase in a standard LT D$_2$ TDS.

Future development for hydrogen storage in 3D graphene may include functionalization with metal atoms or clusters that already proved successful in enhancing the hydrogen storage in flat graphene, such as Pd, Pt, or Ti \cite{Jain2019,Basta_2018,Murata_2018,Takahashi_2016}. Given the complex topography of the graphene structure, issues like metal penetration into the structure, deposition temperature, nanoparticle adhesion, and coverage will be addressed. The combination of the hydrogen storage properties of the 3D graphene and the catalytic effect of the deposited metal adatoms are likely to result in the enhancement of the hydrogen storage performance.

It is important to underline that the porousified surface of an etched SiC wafer can be detached from the bulk crystal. This is a major benefit for applications in the field of hydrogen storage, where the mass of the substrate must be minimized.

3D graphene samples could also result beneficial in the implementation of electrodes in electrolytic cells. In particular, it may find application in the production composites for cathodes in water splitting cells, since a double functionality both as a cathode support and as a storage device would allow the immediate stockpiling of hydrogen into
the electrode as it gets electrically produced.

Hydrogen chemisorption on the 3D graphene nanostructure could also play a role in the promotion of the hydrogen evolution reaction (HER). Indeed, the binding of hydrogen atoms could create new reaction sites for the Heyrovsky reaction, thus allowing HER to also take place away from the immediate
vicinity of the catalytic centres.

TiO$_2$ was proven to promote the photocatalytic dissociation of water \cite{FUJISHIMA_1972}. The drawback of this procedure is the co-evolution of H$_2$ and O$_2$ in the same vessel, resulting in the production of an explosive mixture. A novel nanocomposite of TiO$_2$ particles dispersed in a 3D graphene substrate could catalyse the water splitting reaction while retaining the hydrogen inside its structure, thus lowering the risks associated to the procedure.

In conclusion, three-dimensional graphene structures, conformally grown on the surface of porousified, monocrystalline 4H-SiC, are a promising technology platform for a wide range of possible applications, in particular concerning the field of hydrogen production and storage, as well as catalysis and sensors. The experimental endeavour reported in the present work was able to prove some fundamental properties of the novel material system and provides a guide toward further optimisation with the aim of achieving hydrogen storage capability.

\section{Methods}

\subsection{Porous SiC}

The 4H-SiC wafers used in the present work have a thickness of $350 \ \mu$m and bulk resistivity $0.106$ $\Omega\cdot$cm. The crystals are etched using two successive photo electrochemical steps.

The process of Metal-Assisted Photochemical Etching (MAPCE) consists in the sputter deposition of $300$ nm thick Pt pads on the wafer. The sample is then inserted in a standard electro-chemical cell (AMMT GmbH) containing an electrolytic solution of $0.15$ mol/L Na$_2$S$_2$O$_8$ and $1.31$ mol/L HF.

A $250$ W ES280LL mercury lamp is used as UV source. Pt acts as cathode while
the SiC surface exposed to the solution acts as the anode.
The oxidant is reduced at the cathode, while SiC is oxidised to SiO$_2$ at
the anode and gets dissolved by HF, creating pores at the surface of the substrate \cite{Leitgeb_2016}. The $\mu$m-range porosity of MAPCE acts as the site from which the finer etching process of Photo-Electro Chemical Etching (PECE) takes place. PECE uses a solution of $5.52$ mol/L HF and $1.7$ mol/L ethanol and the same $250$ W ES280LL mercury lamp as UV source.
The sample is placed in-between two compartments of the PECE cell and acts as a separation wall between the two electrodes.
When a bias is applied, SiC is oxidised at the anode, and dissolved by HF, etching pores on the face adjacent
to the electrode with the negative potential \cite{Leitgeb_2016}.

\subsection{Graphene Growth and Characterization}

Every sample entering the UHV system was subjected to a degassing process to remove atmospheric contaminants. The procedure consists in the heating of the sample to 700~$^\circ$C for $10$ hours.

The graphene growth procedure followed in this work is a slight modification of the one described in the work by Veronesi et al.~\cite{Veronesi_2022} and consists in two series of $2.5$ min annealings of the porous SiC substrate at a temperature of 1370~$^\circ$C in the UHV chamber.

\subsection{Hydrogenation}

The hydrogenation procedure followed in this work is the same for every measurement. It consists in the exposure of the sample to a molecular or atomic deuterium atmosphere of pressure $10^{-7}$ mbar for 5 minutes. 
After $5$ minutes, the leak valve feeding hydrogen to the vacuum chamber is closed, and the background pressure is allowed to return to its original value of the order of $10^{-11}$ mbar.

To perform an atomic hydrogenation, the deuterium flow is passed through a hydrogen cracker before entering the chamber, allowing the splitting of the molecules. The instrument used in this work is a Tectra atomic hydrogen source.

The use of deuterium instead of protium is motivated by the fact that the former can be detected with a larger signal to noise ration by the mass spectrometer, since it is less abundant in the residual atmosphere of the vacuum chamber.

\subsection{TDS Measurements}

During a TDS measurement, the sample
and the adsorbed analyte are gradually heated with a linear heating ramp of slope $\beta$, until the adsorbate-substrate bond gets broken
by the thermal excitation and the adsorbed species released.

For this work, TDS measurements were performed in UHV conditions (base pressure $10^{-11}$ mbar). The mass spectrometer used in this work is a SRS RGA, set to register the signal of ions having mass $4$, corresponding to molecular deuterium. Here, $\beta=4 \ ^\circ$Cs$^{-1}$, maintained via the use of a PID temperature feedback, whose parameters were adjusted for each specific TDS measurement and each different sample.

To monitor the desorption process, a mass spectrometer is operated simultaneously to the sample heating, while the sample is placed about 10 cm far from RGA. The spectrometer registers the variation in the partial pressure of target ions having a set mass. The resulting plot of the mass spectrometer signal as a function of the sample temperature is called a TDS spectrum and yields numerous information regarding the desorption process and the substrate-analyte interaction.

After hydrogenation, the sample temperature is maintained constant, therefore in a RT TDS the initial temperature is 30~$^\circ$C, while it is $-160$~$^\circ$C in a LT TDS. Sample cooling is achieved by flowing nitrogen gas, close to liquefaction temperature, in the manipulator cooling system. The sample holder sits on the manipulator and its cooling limit is about $-160$ °C.

Every TDS measuremnt reported in the following was always accompanied by a blank spectrum, in which the same sample was subjected to TDS, without previously exposing it to hydrogenation. The comparison of blank spectra and spectra taken after hydrogenation allows the exclusion of any spectral feature not imputable to hydrogen desorption (an example of blank spectrum is reported in Figure S3 of the SI).

Comparative D$_2$ TDS spectra were also taken on flat epitaxial graphene samples, under identical measurement conditions (see Figures S$4$ and S$5$ in the SI). These measurement confirm the fact that the TDS peaks observed in the following measurements are imputable to hydrogen desorbing from the $3$D sample, and not from spurious effect (such as desorption from the sample holder or chamber walls).
 
All TDS measurements in this work were performed up to a temperature of 700~$^\circ$C, which is the sample degassing temperature, and higher energy chemisorption interactions are not expected.

\subsection{Quantitative analysis}

Desorption is a temperature-activated process. The rate of molecules $N$ detaching from the surface  (i.e. the variation of the sample coverage $\sigma$) is described through the Arrhenius equation \cite{de_Jong_1990}.
\begin{equation}
N(T)=-\frac{d\sigma}{dt}=\nu_n\cdot\sigma^n e^{-\frac{E_d}{RT}},
    \label{Arrhenius}
\end{equation}
where $n$ is the order of the desorption kinetics, $\nu_n$ the rate constant, $E_d$ the desorption activation energy, and $R$ the gas constant.

TDS measurements use a linear heating ramp. Therefore, the temperature is $T=T_0+\beta t$, where $\beta$ is the heating rate and $T_0$ the initial sample temperature. Atomic hydrogen desorption is a first-order
process \cite{Mashoff_2013,Denisov_2001}, and Equation \ref{Arrhenius} can be solved to find the peak temperature $T_p$ \cite{Redhead_1962}:
\begin{equation}
\frac{E_d}{RT_p^2}=\frac{\nu_1}{\beta}e^{-\frac{E_d}{RT_p}}.
    \label{Tpeak}
\end{equation}
By defining $\tau_p$ as the time required by the heating ramp to rise the temperature from $T_0$ to $T_p$, Equation \ref{Tpeak} can be written in the form
\begin{equation}
\frac{E_d}{k_BT_p}=A\tau_p e^{-\frac{E_d}{k_BT_p}},
    \label{tau}
\end{equation}
where $A$ is the \textit{Arrhenius constant}, with typical value of $10^{13}$ s$^{-1}$, and $k_B$ is the Boltzmann constant. The analysis of a TDS spectrum yields both $\tau_p$ and $T_p$, therefore, Equation \ref{tau} can be solved to obtain $E_d$.

\bibliography{Hstor3Dgr}

\end{document}